\documentclass[reprint,amsmath,amssymb,pra]{revtex4-1}
\usepackage{soul}
\usepackage{graphicx}
\usepackage{dcolumn}
\usepackage{bm}
\begin{document}

\title{Nonparaxial electromagnetic Bragg scattering in periodic media with $\mathcal{PT}$ symmetry}

\author{P. A. Brand\~ao}
\email{paulo.brandao@fis.ufal.br}
\author{J. H. Nascimento}
\email{henrique.nascimento@fis.ufal.br}
\author{S. B. Cavalcanti}%
\email{sbessa@gmail.com}
\affiliation{Universidade Federal de Alagoas, Cidade Universit\'{a}ria, Macei\'{o}-AL, 57072-970, Brazil}

\date{\today}

\begin{abstract}

The evolution of a pair of resonant 
Bragg modes 
through a medium characterized by a 
complex one-dimensional $\mathcal{PT}$-symmetric periodic 
permittivity is thoroughly 
investigated. Analytic solutions of 
Maxwell's equations are 
derived beyond the paraxial 
approximation to investigate the periodic 
energy exchange that occurs between 
the Bragg modes for the Hermitian lattices as well as 
for complex lattices. 
Three regimes defined 
by the symmetry breaking point are
discussed: below it, above it and 
at it.  These regimes are determined by 
the existence 
of
four real eigenvalues in the symmetric phase, 
which collide and coalesce into a pair
at the breaking point. 
Above the critical value each member of the pair 
bifurcates
into a pair of complex values.
Therefore, the complex lattice reveals a variety 
of wave
dynamics depending on the gain/loss balance. In all 
regimes of the transition the signature of $\mathcal{PT}$-symmetric systems is present,
as the evolution is always nonreciprocal 
and unidirectional.

\end{abstract}

\maketitle


\section{Introduction}

Quantum mechanics, originally formulated in terms of Hermitian physical observables, has been extended in the past twenty years to include complex operators invariant under parity  and temporal  symmetry transformations. These  $\mathcal{PT}$-symmetric operators \cite{bender,bender01,bender02,bender03,bender04} also have a real spectrum under certain conditions as the $\mathcal{PT}$ symmetry requirement alone does not guarantee the existence of a real spectrum. Actually, the Hamiltonian $H(b)$ contains a free parameter $b$ that may be increased up to a critical value, above which the system undergoes a symmetry breaking phase transition. Below the critical value, all eigenvalues are real and, due to the symmetry $\mathcal{C}$, a properly defined $\mathcal{CPT}$ inner product may be defined to achieve unitary evolution \cite{bender05}. Furthermore, the $\mathcal{PT}$ symmetry is not broken as the eigenvectors of the $\mathcal{PT}$ operator are simultaneous eigenvectors of the Hamiltonian. On the other hand, above the critical value, the phase of the system is said to be spontaneously broken and the Hamiltonian no longer shares a common set of eigenvectors with the $\mathcal{PT}$ operator. Thus, the real eigenvalues give way to complex-conjugated pairs of eigenvalues. 

These $\mathcal{PT}$-symmetric Hamiltonians are quite suitable to describe optical structures due to the similarity between the Schr\"odinger equation and the paraxial wave equation. The latter is obtained from Maxwell's relations and describes the propagation of a wave through a medium with balanced gain/loss. Thus, based on the association of the potential function with complex-valued refractive indices, it was shown recently that even non-$\mathcal{PT}$ operators may be used to represent classes of materials with arbitrary gain and loss \cite{yang}. Experimental evidence on systems that simulate $\mathcal{PT}$-symmetric behavior has been carried out in coupled waveguides \cite{makris2008beam,guo2009observation,ruter2010observation}, silicon photonic circuits \cite{feng2011nonreciprocal}, superconducting wires \cite{rubinstein2007bifurcation} and even in classical mechanical systems \cite{bender2013observation}, to cite a few. 

The present work is mainly concerned with periodic photonic lattices. In this context, the propagation of wide monochromatic Gaussian beams has been studied in a $\mathcal{PT}$-symmetric periodic structure reporting double refraction and power oscillations which are not present in conservative Hermitian systems \cite{makris}. Also, 
it has been demonstrated that the evolution of a paraxial wave under the two-beam approximation
diffracted by a $\mathcal{PT}$-symmetric optical lattice obeys a different sum rule for the intensity of the wave as compared to the Hermitian case \cite{Berry}. Actually, the optical properties of photonic lattices have also been associated with X-rays in crystals, in the sense that these Bragg oscillations have been previously identified with the Pendell\"osung effect in photonic crystals, such as in the 2D case \cite{Mocella}, and in opal 3D photonic crystals \cite{Agio}. Furthermore, experiments on the properties of microwave diffraction in periodic structures have been reported  in 2D artificial dielectric media \cite{Leo}, and in the optical regime in volume holographic gratings \cite{Calvo}. Bragg oscillations have been reported in $\mathcal{PT}$-symmetric photonic lattices \cite{Feynman Jr.17}. 

However, the analyses developed in these works rely on the paraxial approximation limiting its applications to the particular case of optical systems for which all spatial scales are much larger than the wavelength of light. In a photonic crystal this condition is not satisfied, as the wavelength is of the order of the periodicity of the structure \cite{Ramakrishna}, and therefore a more accurate nonparaxial approach is highly desirable to unravel the meaning and accuracy of the paraxial Pendell\"osung effect. Recently, a nonparaxial approach to investigate the Pendell\"osung effect in a finite $\mathcal{PT}$-symmetric photonic lattice, via another route, has reported asymmetric changes in the intensity profile of the field as well as transparency changes \cite{bushuev}. It should also be mentioned that the propagation of electromagnetic waves through localized and periodic media under a nonparaxial regime has been recently considered by some authors \cite{malomed,jones}.

In this work, a theoretical analytic investigation on the propagation of a wide beam through a transversal periodic photonic lattice described by a 
$\mathcal{PT}$-symmetric
electric permittivity is carried out. Considering 
that our study is focused on the Bragg incidence angle, we use a two-waves model to investigate the power 
exchange that occurs between a pair of resonant 
Bragg modes, within three scenarios: (i) Hermitian Bragg scattering (optical Pendell\"osung effect), (ii) $\mathcal{PT}$-symmetric 
Bragg scattering, below and above the symmetry breaking point, and (iii) Bragg scattering at the symmetry 
breaking point. To this end, next section is 
devoted to a general $n$-waves treatment while 
in section III we focus on the simplified two-waves 
version. Sections IV, V, and VI deal with the 
particular cases described in (i), (ii), 
and (iii) respectively. 

\section{General theory}

The propagation of monochromatic electromagnetic 
fields, $\mathbf{E}(\mathbf{r},t) = \text{Re}
[\mathbf{E}(\mathbf{r})\exp(-i\omega t)]$ 
and $\mathbf{H}(\mathbf{r},t) = \text{Re}[\mathbf{H}(\mathbf{r})
\exp(-i\omega t)]$, 
with angular frequency $\omega$, is governed by Maxwell's 
equations (in SI units),
\begin{align}
    \label{pair1}
    \begin{split}
    \nabla\times\mathbf{H}(\mathbf{r}) &= -i\omega\varepsilon_{\text{f}}\varepsilon(\mathbf{r})\mathbf{E}(\mathbf{r}),
    \\
    \nabla\times\mathbf{E}(\mathbf{r}) &= i\omega\mu_{\text{f}}\mathbf{H}(\mathbf{r}),
    \end{split}
\end{align}
where $\varepsilon(\mathbf{r}) = 1 + \chi(\mathbf{r})$ is the space-dependent isotropic electric 
permittivity with $\chi$ being the linear electric susceptibility; $\varepsilon_{\text{f}}$ and $\mu_{\text{f}}$ represent the free-space 
electric permittivity and magnetic permeability, respectively. Since our objective is to study Bragg resonance induced effects, we assume the dielectric function to be a one-dimensional periodic function of space, $\varepsilon(x+a) = \varepsilon(x)$, with period $a$, which may be written as a Fourier series 
\begin{equation}
    \label{eps}
    \varepsilon(x) = \sum_{l \in \mathbb{Z}}\varepsilon_{l}e^{ig_{l}x},
\end{equation}
with $\varepsilon_{l}$ being the Fourier coefficient of the permittivity and $g_{l} = 2\pi l/a$ the one-dimensional reciprocal lattice vector. The first Brillouin zone of the lattice is thus located in the $k$-space domain $-\pi/a \leq k_{x} \leq \pi/a$. After isolating the electric field from the system of equations \eqref{pair1}, we obtain the differential equation satisfied by the electric field alone:
\begin{equation}
    \label{electric}
    \nabla\times\nabla\times\mathbf{E}(\mathbf{r}) = \frac{\omega^{2}}{c^{2}}\varepsilon(x)\mathbf{E}(\mathbf{r}),
\end{equation}
with $c=1/\sqrt{\varepsilon_{\text{f}}\mu_{\text{f}}}$ the vacuum speed of light. Next, we consider the electric field to be polarized in the $y$ direction (TE polarization mode) and propagating in the $(x,z)$ plane in such a way that its wavevector component parallel to the lattice variation is coupled to the edges of the Brillouin zones, i.e., it is Bragg-resonant with the structure
\begin{equation}
    \label{efieldspec}
    \mathbf{E}(x,z) = \hat{y}\sum_{n \in \mathbb{Z}}\psi_{n}(z)e^{ \frac{in\pi x}{a} },
\end{equation}
where $\psi_{n}(z)$ is the spectral amplitude of the electric field at a distance $z$ and $\mathbb{Z}$ is the set of integers. The magnetic field is derived from the electric field and is given by
\begin{equation}
    \mathbf{H}(x,z) = \frac{1}{i\omega\mu_{\text{f}}}\sum_{n\in\mathbb{Z}}\left[ \hat{z}\psi_{n}(z)\frac{i n \pi}{a} - \hat{x}\frac{d\psi_{n}(z)}{dz}\right]e^{ \frac{i n \pi x}{a}},
\end{equation}
where the divergence condition $\nabla\cdot\mathbf{H} = 0$ is satisfied, as can be easily verified. Notice that the choices for the polarization and the direction of the material periodicity are consistent with the constraint $\nabla\cdot\mathbf{D} = 0$ where $\mathbf{D} = \varepsilon_{\text{f}}\varepsilon(x)\mathbf{E}$ is the electric displacement vector. By substituting Equations \eqref{eps} and \eqref{efieldspec} into Equation \eqref{electric} we arrive at the following set of coupled linear second-order ordinary differential equations for the evolution of the spectral amplitudes $\psi_{n}(z)$
\begin{equation}
    \label{evolutionA}
    \frac{d^{2}\psi_{n}}{dz^{2}} =  \frac{\pi^{2}n^{2}}{a^{2}}\psi_{n} - \frac{\omega^{2}}{c^{2}}\sum_{l \in \mathbb{Z}}\varepsilon_{l}\psi_{n-2l}.
\end{equation}
Usually, in a given periodic function $\varepsilon(x)$ its Fourier coefficients $\varepsilon_{l}$ slowly decay as $l$ increases. This happens when the medium properties undergo abrupt changes as occurs 
in a stack of two different materials $ABABAB...AB$, for example. 
This is because higher frequencies are necessary in the sum \eqref{eps} to take into account the extremely rapid variations of the permittivity. However, in some very tractable cases, only three of these coefficients are nonzero. We will be mainly interested in the following $\mathcal{PT}$-symmetric permittivity
\begin{equation}
    \label{epsilon}
    \varepsilon(x) = \varepsilon_{0} + \varepsilon_{R}\cos\left( \frac{2\pi x}{a} \right) + i\varepsilon_{I}\sin\left( \frac{2\pi x}{a} \right),
\end{equation}
where $\varepsilon_{0}$, $\varepsilon_{R}$ and $\varepsilon_{I}$ are real positive numbers. It is clear from Equation \eqref{epsilon} that the only nonzero Fourier coefficients in the expansion \eqref{eps} are $\varepsilon_{0}$ and $\varepsilon_{\pm1}$ and are given in terms of $\varepsilon_{R}$ and $\varepsilon_{I}$ by $\varepsilon_{\pm1} = (\varepsilon_{R}\pm\varepsilon_{I})/2$. It should be pointed out that, since our interest is to describe the most common optical systems available, we require the real part of the relative permittivity to be greater than 1. For this to be satisfied, one must have $\varepsilon_{0} > 1 + \varepsilon_{R}$. With only three Fourier coefficients for the permittivity function, Equation \eqref{evolutionA} may now be rewritten as
\begin{equation}
    \label{principal}
    \frac{d^{2}\psi_{n}}{dz^{2}} = \alpha_{n}\psi_{n} - \varepsilon_{1}\psi_{n-2} - \varepsilon_{-1}\psi_{n+2},
\end{equation}
where $\alpha_{n} =  \pi^{2}n^{2}/a^{2} - \varepsilon_{0}$ and we have normalized $z\rightarrow \omega z/c$ and $a\rightarrow \omega a/c$. Equation \eqref{principal} is one of the main results of this paper and we now proceed to study its solutions in more detail.

\section{Two-waves model}

\begin{figure}
\includegraphics[width=0.5
\textwidth]{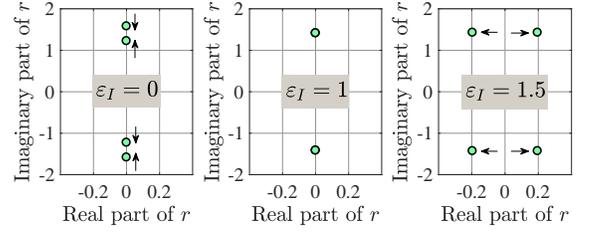}
\centering
\caption{Real and imaginary parts of the eigenvalues below 
(left panel), at (middle panel), and above the critical point. There
are four eigenvalues before they collide at the critical point where they merge 
into two eigenvalues, and give way to a complex conjugated pair. In this figure, $\varepsilon_{R} = 1$ and $\varepsilon_{0} = 3$.}
\label{complexplane}
\end{figure}

Let us consider that only two spectral modes, $\psi_{-1}(z)$ and $\psi_{1}(z)$, are coupled during propagation. If this is the case, Equation \eqref{principal} gives
\begin{align}
    \label{pair2}
    \begin{split}
    \frac{d^{2}\psi_{1}}{dz^{2}} = \alpha\psi_{1} - \varepsilon_{1}\psi_{-1},
    \\
    \frac{d^{2}\psi_{-1}}{dz^{2}} = \alpha\psi_{-1} - \varepsilon_{-1}\psi_{1},
    \end{split}
\end{align}
where $\alpha = \alpha_{\pm1} = 1 - \varepsilon_{0}$ and we have chosen $a = \pi$ without loss of generality. We solve this system by first writing it in standard form $(d/dz)\mathbf{\Psi} = \mathbf{A}\cdot\mathbf{\Psi}$. Explicitly, the matrices $\mathbf{\Psi}$ and $\mathbf{A}$ are given by
\begin{equation}
\label{system2}
\frac{d}{dz}
    \begin{bmatrix}
    \psi_{-1} \\
    \phi_{-1} \\
    \psi_{1} \\
    \phi_{1} 
    \end{bmatrix}
    =
    \begin{bmatrix}
    0 & 1 & 0 & 0\\
    \alpha & 0 & -\varepsilon_{-1} & 0 \\
    0 & 0 & 0 & 1 \\
    -\varepsilon_{1} & 0 & \alpha & 0 
    \end{bmatrix}
    \begin{bmatrix}
    \psi_{-1} \\
    \phi_{-1} \\
    \psi_{1} \\
    \phi_{1} 
    \end{bmatrix},
\end{equation}
where $\phi_{\pm1} = d\psi_{\pm1}/dz$. We now proceed in the standard way of calculating its eigenvalues $r_{j}$ by using $\text{det}(r\mathbf{I}-\mathbf{A})=0$, which gives the fourth-degree characteristic polynomial $r^{4} - 2\alpha r^{2} + (\alpha^{2} - \varepsilon_{1}\varepsilon_{-1}) = 0$. The four eigenvalues are given by
\begin{align}
    \label{eigen}
    \begin{split}
        r_{1} &= -\sqrt{1-\varepsilon_{0} - \frac{1}{2}\sqrt{\varepsilon_{R}^{2}-\varepsilon_{I}^{2}}},\\
        r_{2} &= +\sqrt{1-\varepsilon_{0} - \frac{1}{2}\sqrt{\varepsilon_{R}^{2}-\varepsilon_{I}^{2}}},\\
        r_{3} &= -\sqrt{1-\varepsilon_{0} + \frac{1}{2}\sqrt{\varepsilon_{R}^{2}-\varepsilon_{I}^{2}}},\\
        r_{4} &= +\sqrt{1-\varepsilon_{0} + \frac{1}{2}\sqrt{\varepsilon_{R}^{2}-\varepsilon_{I}^{2}}}.\\
    \end{split}
\end{align}
Figure \ref{complexplane} depicts the dynamics of the eigenvalues \eqref{eigen} as $\varepsilon_{I}$ increases from zero to 1.5 with the values of $\varepsilon_{R} = 1$ and $\varepsilon_{0} = 3$ fixed. If $\varepsilon_{I} = 0$, four distinct pure imaginary 
numbers are shown in part (a) of Figure \ref{complexplane} corresponding to $r = \pm\sqrt{-2\pm1/2}$. On the other hand, when the loss present in the complex 
permittivity is balanced by the gain, $\varepsilon_{I} = \varepsilon_{R} = 1$, two eigenvalues collide at $r = \pm i\sqrt{2}$ giving rise to a system of two eigenvalues 
with multiplicity two. As we will show below, this drastically alters the 
dynamics 
of the propagating waves. When $\varepsilon_{I} > 1$, the four 
eigenvalues become 
distinct once more with nonzero real 
parts. There are, therefore, 
three different behaviors of the field evolution depending 
on the value of $\varepsilon_{I}$ and 
we will study each case in detail.

\begin{figure}[ht]
\includegraphics[width=0.25\textwidth]{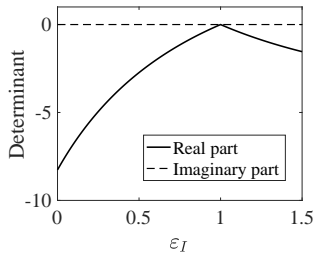}
\centering
\caption{Determinant of the matrix $\mathbf{U}$ as a function of 
$\varepsilon_{I}$, with $\varepsilon_{R} = 1$ and $\varepsilon_{0} = 3$ fixed. 
Since the determinant is different from zero whenever $\varepsilon_{I} \neq 1$, 
the set \eqref{eigenvec} is linearly independent provided we exclude this 
particular value of $\varepsilon_{I}$. }
\label{det}
\end{figure}

The eigenvectors corresponding to the eigenvalues \eqref{eigen} are given by
\begin{align}
\label{eigenvec}
\begin{split}
    \mathbf{u}_{1} = 
        \begin{bmatrix}
            \frac{-\sqrt{\varepsilon_{R}^{2}-\varepsilon_{I}^{2}}}{(\varepsilon_{R}+\varepsilon_{I})\sqrt{1-\varepsilon_{0} - \frac{1}{2}\sqrt{\varepsilon_{R}^{2}-\varepsilon_{I}^{2}}}} \\
            \frac{\sqrt{\varepsilon_{R}^{2}-\varepsilon_{I}^{2}}}{\varepsilon_{R}+\varepsilon_{I}} \\
            -\frac{1}{\sqrt{1-\varepsilon_{0} - \frac{1}{2}\sqrt{\varepsilon_{R}^{2}-\varepsilon_{I}^{2}}}} \\
            1\\
        \end{bmatrix},
\\
    \mathbf{u}_{2} = 
        \begin{bmatrix}
            \frac{\sqrt{\varepsilon_{R}^{2}-\varepsilon_{I}^{2}}}{(\varepsilon_{R}+\varepsilon_{I})\sqrt{1-\varepsilon_{0} - \frac{1}{2}\sqrt{\varepsilon_{R}^{2}-\varepsilon_{I}^{2}}}} \\
            \frac{\sqrt{\varepsilon_{R}^{2}-\varepsilon_{I}^{2}}}{\varepsilon_{R}+\varepsilon_{I}} \\
            \frac{1}{\sqrt{1-\varepsilon_{0} - \frac{1}{2}\sqrt{\varepsilon_{R}^{2}-\varepsilon_{I}^{2}}}} \\
            1\\
        \end{bmatrix},
\\
    \mathbf{u}_{3} = 
        \begin{bmatrix}
            \frac{\sqrt{\varepsilon_{R}^{2}-\varepsilon_{I}^{2}}}{(\varepsilon_{R}+\varepsilon_{I})\sqrt{1-\varepsilon_{0} + \frac{1}{2}\sqrt{\varepsilon_{R}^{2}-\varepsilon_{I}^{2}}}} \\
            \frac{-\sqrt{\varepsilon_{R}^{2}-\varepsilon_{I}^{2}}}{\varepsilon_{R}+\varepsilon_{I}} \\
            -\frac{1}{\sqrt{1-\varepsilon_{0} + \frac{1}{2}\sqrt{\varepsilon_{R}^{2}-\varepsilon_{I}^{2}}}} \\
            1\\
        \end{bmatrix},
\\
    \mathbf{u}_{4} = 
        \begin{bmatrix}
            \frac{-\sqrt{\varepsilon_{R}^{2}-\varepsilon_{I}^{2}}}{(\varepsilon_{R}+\varepsilon_{I})\sqrt{1-\varepsilon_{0} + \frac{1}{2}\sqrt{\varepsilon_{R}^{2}-\varepsilon_{I}^{2}}}} \\
            \frac{-\sqrt{\varepsilon_{R}^{2}-\varepsilon_{I}^{2}}}{\varepsilon_{R}+\varepsilon_{I}} \\
            \frac{1}{\sqrt{1-\varepsilon_{0} + \frac{1}{2}\sqrt{\varepsilon_{R}^{2}-\varepsilon_{I}^{2}}}} \\
            1\\
        \end{bmatrix},
\end{split}
\end{align}
such that $\mathbf{A}\cdot\mathbf{u}_{j} = r_{j}\mathbf{u}_{j}$ for $j = 1,2,3,4$. To verify that the set of eigenvectors \eqref{eigenvec} is linearly independent we construct the $4\times 4$ matrix $\mathbf{U} = [\mathbf{u}_{1}\hspace{0.1cm}\mathbf{u}_{2}\hspace{0.1cm}\mathbf{u}_{3}\hspace{0.1cm}\mathbf{u}_{4}]$ with the columns formed by the eigenvectors \eqref{eigenvec} and calculate its determinant as a function of $\varepsilon_{I}$ with $\varepsilon_{R} = 1$ and $\varepsilon_{0} = 3$ fixed. Figure \ref{det} shows that for every value of $\varepsilon_{I} \neq 1$ the determinant is nonzero and, therefore, the set \eqref{eigenvec} is linearly independent. The particular case where $\varepsilon_{I} = 1$ will be treated separately later. Note also that this corresponds to the situation in Figure \ref{complexplane} where the eigenvalues collide. The general solution of the system \eqref{system2} is given by \cite{brauer}
\begin{equation}
\label{general}
    \mathbf{\Psi}(z) = c_{1}e^{r_{1}z}\mathbf{u}_{1} + c_{2}e^{r_{2}z}\mathbf{u}_{2} + c_{3}e^{r_{3}z}\mathbf{u}_{3} + c_{4}e^{r_{4}z}\mathbf{u}_{4},
\end{equation}
where $c_{j}$ are arbitrary constants determined by the initial conditions. 
In order to write the $c_{j}$'s in terms of $\psi_{j}(0)$ and 
$\phi_{j}(0)$ it is necessary to inspect Equation \eqref{general} 
at $z = 0$. By using the matrix $\mathbf{U}$ and considering the 
column vector $\mathbf{C} = [c_{1}\hspace{0.1cm}c_{2}\hspace{0.1cm}c_{3}\hspace{0.1cm}c_{4}]^{T}$, 
where $T$ is the transpose operation, one may write $\mathbf{\Psi}(0) = \mathbf{U}\cdot\mathbf{C}$ and, after calculating the 
inverse of $\mathbf{U}$, the constants $c_{j}$'s may be 
written as a function of $\mathbf{\Psi}(0)$: $\mathbf{C} = \mathbf{U}^{-1}\cdot\mathbf{\Psi}(0)$. This will give us the 
general solution for $\varepsilon_{I} \neq \varepsilon_{R}$ 
in terms of the initial conditions. 

\section{Hermitian Bragg oscillations (Optical Pendell\"osung effect)}

For comparison purposes with the paraxial approximation, let us begin considering the simplest case of a medium in the absence of gain or loss, with a real electric permittivity ($\varepsilon_{I} = 0$), so that the lattice is Hermitian. By borrowing a very useful 
nomenclature from two-level atomic systems, we define the population inversion function as the difference between the spectral energy content in modes $\psi_{1}$ and $\psi_{-1}$: $W = |\psi_{1}(z)|^{2}-|\psi_{-1}(z)|^{2}$. We define $W_{1}$ to be the population inversion function whose initial spectral energy is fully concentrated within mode $\psi_{1}$, i.e., $\{\psi_{1}(0),\psi_{-1}(0) \} = \{1,0\}$, and $W_{-1}$ to be the population inversion function whose initial spectral energy is fully within mode $\psi_{-1}$, i.e., $\{\psi_{1}(0), \psi_{-1}(0) \} = \{0,1\}$. After a lengthy but straightforward calculation,
it may be shown that under these circumstances, the population inversion functions are given by
\begin{equation}
W_{\pm1} = \pm\cos(\gamma_{-}z)\cos(\gamma_{+}z),
\end{equation}
which are symmetric, as expected, and 
\begin{equation}
\gamma_{\pm} = \text{Im}\left(\sqrt{1-\varepsilon_{0} \pm \frac{\varepsilon_{R}}{2}}\right).
\end{equation}

\begin{figure}[h]
\includegraphics[width=0.40\textwidth]
{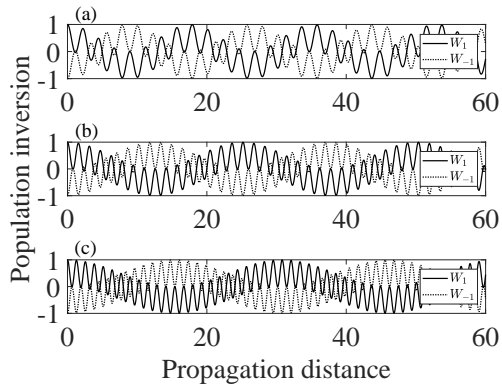}
\centering
\caption{Evolution of the population inversion functions $W_{1}$ (continuous line) and $W_{-1}$ (dotted line) for (a) $\varepsilon_{0} = 3$, (b) $\varepsilon_{0} = 5$ and (c) $\varepsilon_{0} = 7$ with $\varepsilon_{R}= 1$ and $\varepsilon_{I} = 0$.}
\label{population_inversion_e_i_0_fig}
\end{figure}
\begin{figure}[h]
\includegraphics[width=0.40\textwidth]
{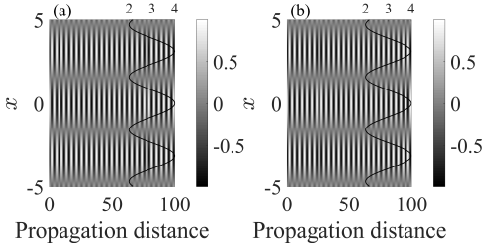}
\caption{Spatial evolution of the 
real part of the electric field 
in nonparaxial electromagnetic 
Hermitian system for (a) $\left\{\psi_{1}(0),\psi_{-1}(0)
\right\}=\left\{1,0\right\}$ 
and (b) $\left\{\psi_{1}(0),\psi_{-1}(0)\right\}=\left\{0,1\right\}$ 
with $\varepsilon_{0} = 3$ and $\varepsilon_{R} = 1$. The black lines in both panels represent the Hermitian electric permittivity with the upper axes indicating its values.}
\label{eletric_field_e_i_0_fig}
\end{figure}

Figure \ref{population_inversion_e_i_0_fig} illustrates the population inversion for three values of $\varepsilon_{0}$ with
$\varepsilon_{R}=1$. First thing we note is that it is a conservative process, as the sum of the power contained in the the resonant Bragg modes is always equals to the input power. Compared to the paraxial case \cite{Feynman Jr.17}, the oscillations exhibit a richer structure. It is clear that two superposed oscillations occur: the envelope, with a much longer spatial period, modulating the phase-like oscillation. The resulting motion is quasi-harmonic in the sense that the envelope amplitude and phase are not exactly constant, although their variation is negligible compared to the faster oscillation. As $\varepsilon_{0}$ increases, the period of the phase oscillations decreases while the modulation cycle increases. This beating behavior is better illustrated by finding the spatial periods of these oscillations and they are given, respectively, by
\begin{equation}
    \Lambda_{\text{p,e}} = \frac{2\pi}{\gamma_{-}\pm\gamma_{+}}
\end{equation}
where $\Lambda_{\text{p}}$ ($\Lambda_{\text{e}}$) stands for the shorter (larger) phase- (envelope-)like oscillation periods. The real part of the electric field $\mathbf{E}(x,z)$ is illustrated in Figure \ref{eletric_field_e_i_0_fig} for $\varepsilon_{0}=3$ and $\varepsilon_{R}=1$. Both initial conditions lead to exactly the same interference pattern, meaning that there is no preferential spatial mode for which energy flows, that is, the energy is shared equally by both modes, and the sum of the power within each mode is equals to the input power, as expected for unitary evolution. The interference pattern obtained in either case, exhibit a transversal variation of contrast following the permittivity function. Minimum values of the permittivity
lead to lower contrast so that in these regions the pattern become dim. 

\section{$\mathcal{PT}$-Symmetric Bragg oscillations with $\varepsilon_{I}\neq1$}

\begin{figure}[h]
\includegraphics[width=0.35\textwidth]
{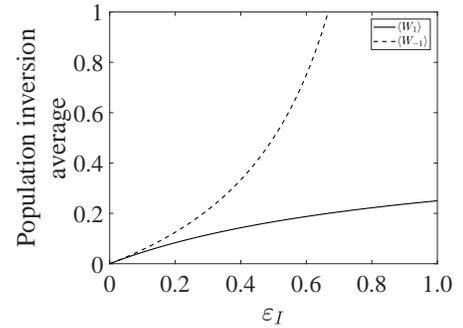}
\centering
\caption{Population inversion averages $\langle W_{1} \rangle$ (continuous line) and $\langle W_{-1} \rangle$ (dashed line). Note the Hermitian case ($\varepsilon_{I}=0$) where both averages are zero).}
\label{average_W_fig}
\end{figure}

In the next two sections we consider systems for which the electric permittivity has a non-zero imaginary part. Let us begin with
the system below the symmetry breaking point, i.e., $0\leq\varepsilon_{I}\leq1$. The population inversion functions in this case are given by
\begin{multline}
W_{-1}=\frac{1-\beta^2}{4\beta^2}\cos^2(\gamma'_{-}z)+
\frac{1-\beta^2}{4\beta^2}\cos^2(\gamma'_{+}z)\\
-\frac{1+\beta^2}{2\beta^2}\cos(\gamma'_{-}z)\cos(\gamma'_{+}z)
\end{multline}
and
\begin{multline}
W_{1}=\frac{1-\beta^2}{4}\cos^2(\gamma'_{-}z)+\frac{1-\beta^2}{4}
\cos^2(\gamma'_{+}z)\\
+\frac{1+\beta^2}{2}\cos(\gamma'_{-}z)\cos(\gamma'_{+}z),
\end{multline}
where
\begin{equation}
\beta=\frac{\sqrt{\varepsilon^2_{R}-\varepsilon^2_{I}}}
{\varepsilon_{R}+\varepsilon_{I}}
\end{equation}
and
\begin{equation}
\gamma'_{\pm}=\text{Im}\left(\sqrt{1-\varepsilon_{0}\pm\frac{1}{2}
\sqrt{\varepsilon^2_{R}-\varepsilon^2_{I}}}\right).
\end{equation}
\begin{figure}[h]
\includegraphics[width=0.40\textwidth]
{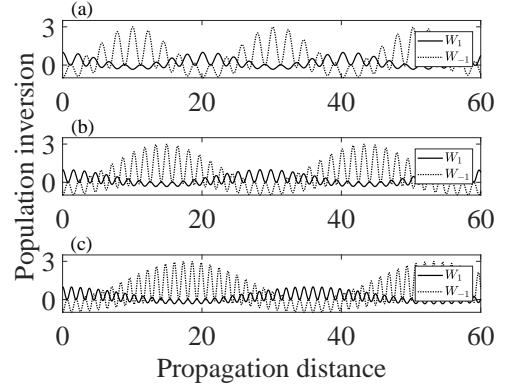}
\centering
\caption{Evolution of the population inversion functions $W_{1}$ (continuous line) and $W_{-1}$ (dotted line) for (a) $\varepsilon_{0} = 3$, (b) $\varepsilon_{0} = 5$ and (c) $\varepsilon_{0} = 7$ with $\varepsilon_{R}= 1$ and $\varepsilon_{I} = 0.5$.}
\label{population_inversion_e_i_less_1_fig}
\end{figure}
Figure \ref{population_inversion_e_i_less_1_fig} illustrates the population inversion functions for three values of $\varepsilon_{0}$ while $\varepsilon_{R}$ is fixed at 1. When $\varepsilon_{I}\neq0$ the symmetric behavior of the population inversion functions exhibited by the Hermitian case no longer exists and this feature is clearly shown in Figure \ref{population_inversion_e_i_less_1_fig}: $W_{-1}$ now oscillates with an amplitude larger than $W_{1}$ during most part of the propagation and, in fact, as $\varepsilon_{0}$ increases it can be seen that the oscillatory behavior of $W_{-1}$ becomes much more evident than that for $W_{1}$, which oscillates with a very small amplitude compared to the amplitude of $W_{-1}$. Phase-like and envelope-like oscillations are still present below the symmetry breaking point, and with the same beating behavior as before, with some differences though. The spatial periods are now given by
\begin{equation}
\Lambda'_{\text{p}}=\frac{2\pi}{2\gamma'_{-}}
\end{equation}
and
\begin{equation}
\Lambda'_{\text{e}}=\frac{2\pi}{\gamma'_{-}-\gamma'_{+}}.
\end{equation}
Comparing to the Hermitian case, the phase-like spatial periods become shorter while the envelope ones become longer. Another feature when the system is below the symmetry breaking point is a non-zero average value for the population inversion functions reflecting the fact that energy is not shared equally by both modes. By noting that the cosine square function and the product of two cosine functions (with different arguments) give average values equal to $1/2$ and zero, respectively, over one complete oscillation, the population inversion averages are given by
\begin{equation}
\langle W_{\pm1} \rangle = \frac{\varepsilon_{I}}
{2(\varepsilon_{R}\pm
\varepsilon_{I})}.
\end{equation}
These averages are illustrated in Figure \ref{average_W_fig} (note a diverging behavior in $W_{-1}$ as $\varepsilon_{I} \rightarrow  1$). Using the same values of $\varepsilon_{R}$ and $\varepsilon_{I}$ as in Figure \ref{population_inversion_e_i_less_1_fig}, one may obtain $\langle W_{1} \rangle \approx 0.16$ and $\langle W_{-1} \rangle \approx 0.5$.
These values for the population inversion averages reflects the fact that there is a privileged mode in the sense that irrespective of the initial condition the spectral energy seems to flow to a particular mode, $\psi_{1}$. More specifically, choosing the spectral energy input at mode $\psi_{1}$, what happens next is that the energy oscillates to and from $\psi_{-1}$ with a quite small amplitude. Furthermore, considering mode $\psi_{-1}$ as the input, the energy transfer to the privileged $\psi_{1}$ mode is enhanced, in the sense that the overall picture is the same, except that now, the amplitude in mode $\psi_{1}$ is much larger than in mode $\psi_{-1}$. This unbalanced energy distribution is clearly seen to diverge at the symmetry breaking point in Figure \ref{average_W_fig}.

\begin{figure}[h]
\includegraphics[width=0.40\textwidth]{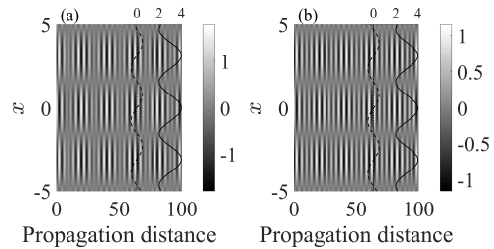}
\caption{Spatial evolution of the real part of the electric field below the symmetry breaking point for (a) $\left\{\psi_{1}(0),\psi_{-1}(0)\right\}= \left\{1,0\right\}$ and (b) $\left\{\psi_{1}(0),\psi_{-1}(0)\right\}= \left\{0,1\right\}$ with $\varepsilon_{0} = 3$, $\varepsilon_{R} = 1$ and $\varepsilon_{I} = 0.5$. The black lines in both panels represent the real (continuous line) and imaginary (dashed line) parts of the electric permittivity and the dotted line separates the regions of gain and loss. The upper axes indicate few values of the permittivity.}
\label{eletric_field_e_i_less_1_fig}
\end{figure}

\begin{figure}[h]
\includegraphics[width=0.40\textwidth]
{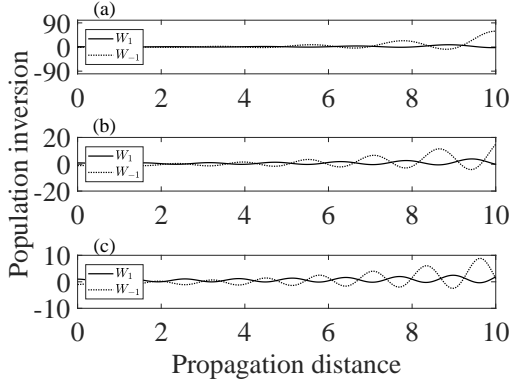}
\centering
\caption{Evolution of the population inversion functions $W_{1}$ (continuous line) and $W_{-1}$ (dotted line) for (a) $\varepsilon_{0} = 3$, (b) $\varepsilon_{0} = 5$ and (c) $\varepsilon_{0} = 7$ with $\varepsilon_{R}= 1$ and $\varepsilon_{I} = 1.5$.}
\label{population_inversion_e_i_greater_1_fig}
\end{figure}

Let us now turn to the spatial evolution 
of the real part of the electric field $\mathbf{E}(x,z)$, which is illustrated 
in Figure \ref{eletric_field_e_i_less_1_fig} 
for $\varepsilon_{0} = 3, \varepsilon_{R} = 1$, and $\varepsilon_{I} = 0.5$. This spatial 
pattern reflects the dynamics demonstrated 
by the population inversion functions 
in Figure 
\ref{population_inversion_e_i_less_1_fig}. Like the Hermitian case, the transverse 
interference pattern is composed by cycles of bright and dark fringes of interference separated by smaller cycles of low contrast fringes. But now, as the maximum value attained by the envelope amplitude of the population inversion function also varies, low contrast regions appears periodically in the propagation direction. These regions correspond to locations where the energy content is minimum and thus the periodical pattern is almost gone, and one may see them in Figure \ref{eletric_field_e_i_less_1_fig} as blurred sections along both the propagation and transverse direction. 
Let us now investigate the system behavior above the symmetry breaking point, that is, for $\varepsilon_{I}>1$. In this case, the eigenvalues of the matrix $\textbf{A}$ have both real and imaginary parts (Figure \ref{complexplane}) revealing an entirely different evolution from the previous ones, found in the last section. The population inversion functions are now given by the expressions
\begin{widetext}
\begin{multline} \label{eq26}
        W_{-1}  = \frac{1-\tilde{\beta}^2}{4\tilde{\beta}^2}[\sinh^{2}(\tilde{\gamma}_{-}z) + \cos^{2}(\tilde{\gamma}_{-}'z)] + \frac{1-\tilde{\beta}^2}{4\tilde{\beta}^2}[\sinh^{2}(\tilde{\gamma}_{+}z) + \cos^{2}(\tilde{\gamma}_{+}'z)] \\
         - \frac{1+\tilde{\beta}^2}{2\tilde{\beta}^2}\left[  \cosh(\tilde{\gamma}_{-}z)\cosh(\tilde{\gamma}_{+}z)\cos(\tilde{\gamma}_{-}'z)\cos(\tilde{\gamma}_{+}'z) + \sinh(\tilde{\gamma}_{-}z)\sinh(\tilde{\gamma}_{+}z)\sin(\tilde{\gamma}_{-}'z)\sin(\tilde{\gamma}_{+}'z) \right]
\end{multline}
and
\begin{multline} \label{eq27}
        W_{1}  = \frac{1-\tilde{\beta}^2}{4}[\sinh^{2}(\tilde{\gamma}_{-}z) + \cos^{2}(\tilde{\gamma}_{-}'z)] + \frac{1-\tilde{\beta}^2}{4}[\sinh^{2}(\tilde{\gamma}_{+}z) + \cos^{2}(\tilde{\gamma}_{+}'z)] \\
         + \frac{1+\tilde{\beta}^2}{2}\left[  \cosh(\tilde{\gamma}_{-}z)\cosh(\tilde{\gamma}_{+}z)\cos(\tilde{\gamma}_{-}'z)\cos(\tilde{\gamma}_{+}'z)  + \sinh(\tilde{\gamma}_{-}z)\sinh(\tilde{\gamma}_{+}z)\sin(\tilde{\gamma}_{-}'z)\sin(\tilde{\gamma}_{+}'z) \right],
\end{multline}
\end{widetext}
where
\begin{equation}
    \tilde{\beta} = \frac{\sqrt{\varepsilon_{I}^2-\varepsilon_{R}^2}}{\varepsilon_{R}+\varepsilon_{I}},
\end{equation}
\begin{equation}
    \tilde{\gamma}_{\pm} = \text{Re}\left( \sqrt{1-\varepsilon_{0} \pm\frac{i}{2}\sqrt{\varepsilon_{I}^2-\varepsilon_{R}^2}} \right)
\end{equation}
and
\begin{equation}
    \tilde{\gamma}_{\pm}' = \text{Im}\left( \sqrt{1-\varepsilon_{0} \pm\frac{i}{2}\sqrt{\varepsilon_{I}^2-\varepsilon_{R}^2}} \right).
\end{equation}
Equations \eqref{eq26} and \eqref{eq27} clearly show an unbounded behavior due to the hyperbolic functions. Figure \ref{population_inversion_e_i_greater_1_fig} illustrates the population inversion functions on propagation, for three different values of $\varepsilon_{0}$ with $\varepsilon_{R}=1$ and $\varepsilon_{I}=1.5$. 
Besides the asymmetry between $W_{1}$ and $W_{-1}$, in this case the oscillations persist, however now they exhibit an unbounded oscillatory behavior. At the very beginning of the propagation the amplitude oscillations are quite small but they increase quickly and grow indefinitely. As before, the oscillation amplitude grows faster by choosing the initial condition $\left\{\psi_{1}(0),\psi_{-1}(0)\right\}=\left\{0,1\right\}$, and also all the energy is quickly transferred to mode $\psi_{1}$, whatever initial condition one considers.
 
\begin{figure}[h]
\includegraphics[width=0.40\textwidth]
{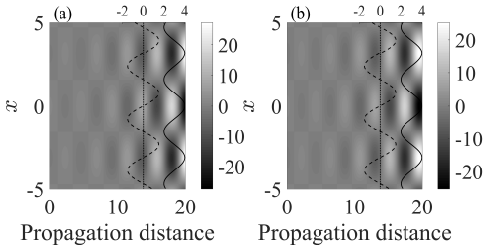}
\caption{Spatial evolution of the real part of the electric field above the symmetry breaking point for (a) $\left\{\psi_{1}(0),\psi_{-1}(0)\right\}=\left\{1,0\right\}$ and (b) $\left\{\psi_{1}(0),\psi_{-1}(0)\right\}=\left\{0,1\right\}$
with $\varepsilon_{0} = 3$, $\varepsilon_{R} = 1$ and $\varepsilon_{I} = 1.5$. The black lines in both panels represent the real (continuous line) and imaginary (dashed line) parts of the electric permittivity and the dotted line separates the regions of gain and loss. The upper axes indicate few values of the permittivity.}
\label{eletric_field_e_i_greater_1_fig}
\end{figure}

Let us now turn to Figure \ref{eletric_field_e_i_greater_1_fig} where  the real part of the electric field $\mathbf{E}(x,z)$ is depicted for $\varepsilon_{0} = 3$, $\varepsilon_{R}=1$ and $\varepsilon_{I}=1.5$. As before, this pattern reflects the population inversion dynamics (Figure \ref{population_inversion_e_i_greater_1_fig}), where at the beginning of propagation, both initial conditions lead to a tiny electric field which grows considerably as propagation continues. The interference fringes begin to appear quite faintly but then, the fringe contrast increases intensely quite rapidly. 

\section{$\mathcal{PT}$-SYMMETRIC BRAGG OSCILLATIONS AT THE 
SYMMETRY BREAKING POINT ($\varepsilon_{I} = 1$)}

As we have seen above, in this case the eigenvectors  become degenerate with  double multiplicity so that the general solution is given by \cite{brauer}

\begin{equation}
\mathbf{\Psi}(z) = \sum_{j=1}^{k}e^{r_{j}z}\left[\sum_{p=0}^{n_{j}-1}
\frac{z^{p}}{p!}\left(\mathbf{A}-r_{j}\mathbf{I}
\right)^{p}\right]\cdot \mathbf{u}_{j},
\end{equation}
where $k$ is the number of eigenvectors $\mathbf{u}_j$, $n_j$ is the multiplicity of the respective eigenvalues and $\mathbf{I}$ is the $4\times4$ identity matrix. It can be shown that
\begin{equation}
\mathbf{\Psi}(z) = \begin{bmatrix} 
\psi_{-1}(0)\cos\left(\tilde{\alpha}^{1/2}z\right) \\
-\tilde{\alpha}^{1/2}\psi_{-1}(0)\sin\left(\tilde{\alpha}^{1/2}z\right) \\
\psi_{1}(0)\cos\left(\tilde{\alpha}^{1/2}z\right)-\frac{\psi_{-1}(0)}{2\tilde{\alpha}^{1/2}}z\sin\left(\tilde{\alpha}^{1/2}z\right) \\
-\frac{\psi_{-1}(0)}{2}z\cos\left(\tilde{\alpha}^{1/2}z\right)-\frac{\psi_{-1}(0)+\tilde{\alpha}\psi_{1}(0)}{\tilde{\alpha}^{1/2}}\sin\left(\tilde{\alpha}^{1/2}z\right)\\
\end{bmatrix},
\end{equation}
where $\tilde{\alpha}=\varepsilon_{0}-1$. The population inversion functions for this case are given by
\begin{equation}
W_{-1}=\frac{1}{4\tilde{\alpha}}z^{2}\sin^{2}(\tilde{\alpha}^{1/2}z)-\cos^{2}(\tilde{\alpha}^{1/2}z)
\end{equation}
and
\begin{equation}
W_{1}=\cos^{2}(\tilde{\alpha}^{1/2}z).
\end{equation}
\begin{figure}[h]
\includegraphics[width=0.40\textwidth]
{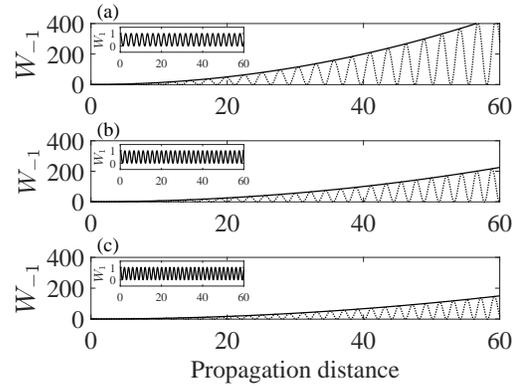}
\centering
\caption{Evolution of the population inversion functions $W_{1}$ (insets) and $W_{-1}$ (dotted line) for (a) $\varepsilon_{0} = 3$, (b) $\varepsilon_{0} = 5$ and (c) $\varepsilon_{0} = 7$ with $\varepsilon_{R}= 1$ and $\varepsilon_{I} = 1$. The continuous line is the parabola $[4 (\varepsilon_0 -1)]^{-1} z^2.$}
\label{population_inversion_e_i_equals_1_fig}
\end{figure}
Figure \ref{population_inversion_e_i_equals_1_fig} illustrates the population inversion functions for three values of $\varepsilon_{0}$ while $\varepsilon_{R}$ is fixed at 1 . In contrast with the paraxial result for which $W_{1}=1$ constant \cite{Feynman Jr.17}, here at the symmetry breaking point we find no occurrence of mode trapping. Actually here, we find that $W_{1}$ oscillates harmonically transferring power to and from efficiently, while the amplitude of the $W_{-1}$ oscillation increases with $z^{2}$. Furthermore, as $\varepsilon_0$ increases, the growth of the amplitude of the oscillations becomes slower. Thus, at the symmetry breaking point, we find a quite asymmetric dynamics comparing $W_{-1}$ and $W_{1}$ and both dynamics are strikingly contrasting with the paraxial Bragg oscillations reported in \cite{Feynman Jr.17}. 

\begin{figure}[h]
\includegraphics[width=0.40\textwidth]{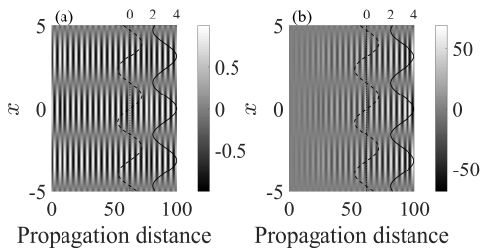}
\caption{Spatial evolution of the real part of the electric field 
at the symmetry breaking 
point for (a) $\left\{\psi_{1}(0),\psi_{-1}(0)\right\}=\left\{1,0\right\}$ 
and (b) $\left\{\psi_{1}(0),\psi_{-1}(0)\right\}=\left\{0,1\right\}$ 
with $\varepsilon_{0} = 3$, $\varepsilon_{R} = 1$ and $\varepsilon_{I} = 1.0$. 
The black lines in both panels represent the real (continuous line) 
and imaginary (dashed line) parts of the electric permittivity and the dotted line separates   
the regions of gain and loss. The upper axes indicate 
few values of the permittivity.}
\label{eletric_field_e_i_equals_1_fig}
\end{figure}

Finally, let us turn to Figure \ref{eletric_field_e_i_equals_1_fig} where we present the 
spatial distribution of the electric field at the symmetry breaking point, with parameters 
$\varepsilon_{0} = 3$, $\varepsilon_{R} = 1$ and $\varepsilon_{I} = 1$. The asymmetric behavior 
of the modes is clear. By choosing mode $\psi_{1}$ as the input, one obtains interference 
fringes of constant contrast. On the other hand choosing mode $\psi_{-1}$ as the input, 
the fringes appear with a linear increasing contrast along the 
propagation direction. 

\section{Conclusions}

Within the frame of Maxwell's equations, we have analyzed the process of energy transfer between a pair of resonant Bragg non-paraxial modes in $\mathcal{PT}$-symmetric photonic lattices. To this end we have used a simple two-waves analytic model to solve the wave equation, and so obtain the mode dynamics for the Hermitian lattice ($\varepsilon_I=0$) as well as for the complex one ($\varepsilon_I\neq 0$). In the Hermitian lattice, the population inversion functions exhibit a beat-like evolution profile, which is reflected in the diffraction pattern of the electric field, and the sum of the energy contained in each mode is always equals to the incident power. Furthermore, there is no asymmetry in the mode dynamics: one may choose mode $\psi_{1}$ or $\psi_{-1}$ as the input, to observe the same dynamics. This is a contrasting feature with the complex lattices which exhibit severe nonreciprocal unidirectional propagation. Furthermore, the sum of the mode energies is not equal to the input energy. Also, in this case a symmetry breaking phase transition occurs for a critical value of the imaginary part of the electric permittivity, at which four real eigenvalues collide and merge into a pair. Above the critical value each member of the pair gives way to a pair of complex values. Therefore, one finds dramatically different dynamics in each case. When the $\mathcal{PT}$ symmetry is not broken, one obtains population inversion oscillations with variable amplitude. At the critical point where the eigenvalues are degenerated, the beat-like pattern of the population inversion disappears and depending on which mode one chooses for the input, one may find harmonic oscillations ($\psi_{1}$) or oscillations whose envelope may grow indefinitely ($\psi_{-1}$). 

Therefore, our results have shown that within the framework of non-Hermitian photonic lattices, which are well suited to describe periodic optical systems with balanced gain/loss profile, a variety of different beam dynamics is unveiled. Apart from their fundamental interest, our results suggest new routes for modal tailoring and control based on $\mathcal{PT}$-symmetric photonic lattices. 

\section*{Acknowledgements}

The authors would like to acknowledge the Brazilian Agencies CNPq,
CAPES and FAPEAL for financial support.

\end{document}